\documentclass[12pt]{article}
\usepackage{amssymb}
\def\aa{{\cal A}}
\def\bb{{\cal B}}

\def\dd{{\cal D}}
\def\ee{{\cal E}}

\def\mm{{\cal M}}

\def\pp{{\cal P}}

\def\zz{{\cal Z}}

\def\nnn{\mathbb N}
\def\ccc{\mathbb C}

\def\rrr{\mathbb R}
\def\zzz{\mathbb Z}

\def\lp{\left(}
\def\rp{\right)}

\def\ov{\overline}
\def\ot{\otimes}

\def\bbb{\begin{equation}}
\def\eee{\end{equation}}
\def\bbbb{\begin{eqnarray}}
\def\eeee{\end{eqnarray}}
\newcommand{\nn}{\nonumber}
\def\eqn#1{(\ref{#1})}

\newtheorem{proposition}{Proposition}[section] 
\newcommand{\bprop}{\medskip\begin{proposition} \it}
\newcommand{\eprop}{\end{proposition} \hfill \medskip \\} 
\newcommand\IC{{\mathbb C}}

\newcommand\IZ{{\mathbb Z}}
\newcommand\IR{{\mathbb R}}

\def\inf1{{\cal L}^{(1,\infty)}}

\def\pa{\partial}

\def\bra#1{\left\langle #1\right|}
\def\ket#1{\left| #1\right\rangle}
\def\hs#1#2{\left\langle #1,#2\right\rangle}  %
\def\ca{{\cal A}}

\def\ce{{\cal E}}

\def\cs{{\cal S}}

\def\raw{\rightarrow}

\begin{document}
\setcounter{page}{0}
\thispagestyle{empty}
\begin{flushright} SISSA 158/99/FM\\ DSM--QM464\\
\hfill March 2000 \\
\end{flushright}

\vspace{.5cm}
\begin{center}{\Large \bf Some Properties of Non-linear $\sigma$-Models \\ ~ \\ in
Noncommutative Geometry} \end{center} \vspace{1cm}
\centerline{\large Ludwik Dabrowski $^1$, Thomas Krajewski $^{1,2}$, Giovanni Landi
$^{2,3}$}
\vspace{5mm}
\begin{center}
$^1$ {\it Scuola Internazionale Superiore di Studi Avanzati,\\ Via Beirut 2-4,
I-34014, Trieste, Italy \\ \vspace{5mm} $^2$ Dipartimento di Scienze Matematiche,
Universit\`a di Trieste \\ P.le Europa 1, I-34127, Trieste, Italy \\ \vspace{5mm}
$^3$ INFN, Sezione di Napoli, Napoli, Italy.} 
\end{center} 
\vspace{1cm}
\begin{abstract} We introduce non-linear $\sigma$-models in the framework of
noncommutative geometry with special emphasis on models defined on the
noncommutative torus. We choose as target spaces the two point space and the circle
and illustrate some characteristic features of the corresponding $\sigma$-models.
In particular we construct a $\sigma$-model instanton with topological
charge equal to $1$. We also define and investigate some properties of a
noncommutative analogue of the Wess-Zumino-Witten model.
\end{abstract}
\vspace{2cm}
\begin{center} Talk presented by T. Krajewski at the Euroconference\\ ``Hopf
algebras and noncommutative geometry in field theory and particle physics'' Torino,
Villa Gualino, September 1999.
\end{center}
\vfill\eject

\section{Introduction}

One could say that the aim of noncommutative geometry is to generalize geometrical
tools to ``spaces whose coordinates fail to commute'' \cite{book}, (see also
\cite{madore},
\cite{landi}, \cite{varily}). One way to implement this program is to start with a
given geometrical theory involving sets $X$ endowed with additional structures and
formulate them algebraically by using suitable subalgebras of the algebra of
complex valued functions over $X$. Then one extends parts of the theory to
noncommutative algebras, which are thought of as functions over ``noncommutative
spaces''. Although much of the construction takes place at the algebraic level, it
is necessary, in order to use the powerful machinery of functional analysis, to
represent these algebras as operators on a Hilbert space. Accordingly, this can be
seen from a physicist's point of view, as analogous to quantum mechanics: one
trades the commutative algebra of functions over phase space for a noncommutative
algebra of operators acting on a Hilbert space.
Most of the geometrical ideas of classical mathematics can be ``quantized''. For
example, topology can be formulated in terms of C$^{*}$-algebras, commutative
C$^{*}$-algebras corresponding to locally compact spaces. Thus, noncommutative ones
are referred to as ``continuous functions over noncommutative locally compact
spaces''. In the compact case (i.e. when the algebra is unital), one can further
define noncommutative vector bundles as finitely generated projective modules over
a given unital C$^{*}$-algebra which plays the role of functions over the base
space. When this algebra is commutative, Serre-Swan's theorem asserts that these
modules correspond to module of sections of vector bundles. Furthermore, methods of
differential topology are also available within the realm of cyclic and Hochschild
cohomology and this leads, via the coupling of the former with K-theory, to
quantities that are stable under deformation and that generalize topological
invariants, like, for instance, winding numbers and topological charges.

Noncommutative geometry has already proved to be useful in understanding various
physical phenomena, like the integral quantum Hall effect \cite{bellisard} or the
classical aspect of the Higgs sector of the standard model (see \cite{schucker} for
a review). Recent developments \cite{CDS} and \cite{SW} also indicate that it is
helpful in string theory. These last developments involve Yang-Mills fields defined
on noncommutative spaces that fit into a broad formalism for gauge fields in
noncommutative geometry which allows one to define connections, their curvature
or the associated Yang-Mills action while preserving most of their classical
aspects. For instance, one can prove a topological bound for the Yang-Mills action
in dimension 4 \cite{book}. Also, one can construct a Chern-Simons type theory and
interpret its behavior under large gauge transformations as a coupling between
cyclic cohomology and K-theory \cite{che}.

In this report, we will be interested in constructing analogues of two dimensional
non-linear $\sigma$-models within the noncommutative world. Since these models
usually exhibit a very rich and easily accessible geometrical structure, we expect
their noncommutative counterparts to be an ideal playground for a probe into the
interplay between noncommutative geometry and field theory. This we shall try to
exemplify by means of three different models: a continuous analogue of the Ising
model which admits instantonic solutions, the analogue of the principal chiral
model together with its infinite number of conserved currents and the
noncommutative Wess-Zumino-Witten model together with its modified conformal
invariance.

All ideas will be presented in a rather sketchy form and we refer to \cite{non1}
(for fields with values in finite spaces) and \cite{non2} (for $S^{1}$-valued
fields) for a detailed account. 

\section{General Aspects}

In ordinary field theory, non-linear $\sigma$-models (see \cite{ICTP} for a review)
are field theories whose configuration space consists of maps $X$ from  a
Riemannian manifold $\Sigma$ with metric $g$, which we assume to be compact and
orientable, to an other Riemannian manifold $\mm$ whose metric we denoted by $G$.
In the physics literature, these manifolds are called source and target space
respectively. By using local coordinates, the action functional is defined as
\bbb\label{sigmaaction} S[X] = {1 \over 2\pi} \int_{\Sigma}
\sqrt{g} ~g^{\mu\nu}\,G_{ij}(X)\partial_{\mu}X^{i}\,\partial_{\nu}X^{j}, \eee where
as usual $g=\det g_{\mu\nu}$ and $g^{\mu\nu}$ is the inverse of $g_{\mu\nu}$. When
$\Sigma$ is two dimensional, the action $S$ is conformally invariant, since a
rescaling of the metric $g\;\rightarrow\;ge^{\sigma}$, with $\sigma$ being any map
from $\Sigma$ to $\rrr$, leaves it invariant. Accordingly, the action only depends
on the conformal class of the metric and may be rewritten using a complex structure
on $\Sigma$ as
\bbb S[X] = {i \over \pi}\int_{\Sigma}\,G_{ij}(X)\,\partial
X^{i}\wedge\ov{\partial}X^{j} \eee where $\partial=\partial_{z}dz$ and
$\ov{\partial}=\partial_{\ov{z}}d\ov{z}$, $z$ being a suitable local complex
coordinate.

Different choices of the source and target spaces lead to different field theories,
some of them playing a major role in physics. Especially interesting are their
applications to statistical field theory, and (supplemented by some
supersymmetries) they are the basic building blocks of superstring theories.

{}From the mathematical point of view, the stationary points of the action
functional \eqn{sigmaaction} are harmonic maps from $\Sigma$ to $\mm$ and describe
the extremal surfaces embedded in $\mm$. Thus, a noncommutative generalization of
the action functional of the non-linear $\sigma$-model should yield, as stationary
points, noncommutative analogues of harmonic maps. 

To generalize such a construction to the noncommutative case, we must dualize the
previous picture and reformulate it in terms of the $*$-algebras $\aa$ and $\bb$ of
complex valued smooth functions defined respectively on $\Sigma$ and $\mm$.
Embeddings $X$ of $\Sigma$ into $\mm$ are in one to one correspondence with
$*$-algebra morphisms $\pi$ from $\bb$ to $\aa$, the correspondence being simply
$\pi_{X}(f)=f\circ X$. Since this makes perfectly sense in the noncommutative case,
we define our configuration space, for fixed, not necessarily commutative algebras
$\aa$ and $\bb$, as the space of all $*$-algebra morphisms from $\bb$ to $\aa$.

The construction of the action functional is more tricky since it involves
noncommutative generalizations of the conformal and Riemannian geometries.
Following an idea of Connes \cite{book} and \cite{newconnes}, the former can be
understood within the framework of positive Hochschild cohomology. Without entering
into details, one observes that, in the commutative case, the trilinear map on
$\aa^{3}$ defined by
\bbb\label{trilinear}
\phi(f_{0},f_{1},f_{2})= {i \over \pi} \int_{\Sigma}f_{0}\partial
f_{1}\wedge\ov{\partial}f_{2}
\eee is an extremal element of the space of positive Hochschild cocycles that
belongs to the cohomology class of the cyclic cocycle $\psi$ defined by
\bbb\label{trilinearbis}
\psi(f_{0},f_{1},f_{2})= {i \over 2 \pi}\int_{\Sigma}f_{0} df_{1} \wedge df_{2}.
\eee Again, we refer to
\cite{book} for the general definitions and we simply notice that
\eqn{trilinear}-\eqn{trilinearbis} still make sense for a general noncommutative
algebra $\aa$.

Roughly speaking, one can say that
$\psi$ allows to integrate 2-forms in dimension 2, \bbb {i \over 2\pi} \int
a_{0}da_{1}da_{2}=\psi(a_{0},a_{1},a_{2}) \eee so that it is a metric independent
object, whereas $\phi$ defines a suitable scalar product \bbb
\langle a_{0}da_{1},b_{0}db_{1}
\rangle=\phi(b_{0}^{*}a_{0},a_{1},b_{1}^{*}) \eee on the space of 1-forms and thus
depends on the conformal class of the metric. Furthermore, this scalar product is
positive and invariant with respect to the action of the unitary elements of $\aa$
on 1-forms, and its relation to the cyclic cocylic $\psi$ allows to prove various
inequalities involving topological quantities.

Having such a cocycle $\phi$, it is natural to compose it with a morphism
$\pi:\;\bb\rightarrow\aa$ in order to obtain a positive cocycle on $\bb$ defined by
$\phi_{\pi}=\phi\circ (\pi\ot\pi\ot\pi)$. Since our goal is to build an action
functional, which assigns a number to any morphism $\pi$, we have to evaluate the
previous cocycle on a suitably chosen element of $\bb^{\ot 3}$. Such an element is
provided by the noncommutative analogue of the metric on the target, which we take
simply as a positive element $G=\sum_{i}b_{0}^{i}\delta b_{1}^{i}\delta b_{2}^{i}$
of the space of universal 2-forms $\Omega^{2}(\bb)$. Thus \bbb S[\pi]=\phi_{\pi}(G)
\eee is well defined and positive and we take it as a noncommutative analogue of
the action functional of the non linear $\sigma$-model. Of course we consider $\pi$
as the dynamical variable (the embedding) whereas $\phi$ (the conformal structure
on the source) and $G$ (the metric on the target) are background structures that
have been fixed.

As an alternative, one could consider that only the metric $G$ on the target is a
background field, since the morphism $\pi:\;\bb\rightarrow\aa$ allows to define the
induced metric $\pi_{*}G$ on the source as \bbb
\pi_{*}G=\mathop{\sum}\limits_{i}\pi(b_{0}^{i})\delta\pi( b_{1}^{i})\delta\pi(
b_{2}^{i}),
\eee which is obviously a positive universal 2-form on $\aa$. To such an object one
can associate, by means of a variational problem (see \cite{book} and
\cite{newconnes}), a positive Hochschild cocycle that stands for the conformal
class of the induced metric. As a result, the critical points of the corresponding
$\sigma$-model describe ``minimally embedded surfaces'' in the noncommutative space
associated with $\bb$.

A scrupulous reader may be puzzled by such a formal and sketchy construction.
However, in what follows we will mainly work out examples involving the
noncommutative torus and only consider a fixed $\phi$ and fixed metrics on the two
target space we will consider (the circle and the two point space). Accordingly,
$\phi$ and $G$ could be replaced by their explicit expressions. Nevertheless, we
think that it may be useful to have a general setting. In particular, one easily
reconstructs ordinary $\sigma$-models with suitable choices of $\phi$ and $G$.

\section{Two points as a target space}

\subsection{A General Construction}

The simplest example of a target space one can think of is that of a finite space
made of two points, like in the Ising model. Of course, any continuous map from a
connected surface to a discrete space is constant and the resulting (commutative)
theory would be trivial. However, this is no longer true if the source is a
noncommutative space and one has, in general, lots of such maps (i.e. algebra
morphisms).

Let us first notice that the algebra $\bb=\ccc^{2}$ of functions over a two point
space is the unital algebra generated by a hermitian projection $e, e^2 = e^* = e$.
Thus, any
$*$-algebra morphism $\pi$ from $\bb$ to $\aa$ is given by a hermitian projection
$p=\pi(e)$ in $\aa$. Choosing the metric $G=\delta e\delta e$ on the space of two
points, the action functional simply reads \bbb S[p]=\phi(1,p,p), \eee where $\phi$
is a given Hochschild cocycle standing for the conformal structure. Of course, one
could choose other metrics on the two points space, but
$G$ is more interesting since it will lead to a topological bound for the
corresponding action. We shall prove this fact for the noncommutative torus, but
the procedure is general and only uses the idea of positivity in Hochschild
cohomology.

\subsection{The noncommutative two torus as a source space} 

For the sake of completeness we recall the very basic aspects of the noncommutative
torus and refer the reader to \cite{rieffel} for a thorough survey. The algebra
$\ca_\theta$ of smooth functions on the noncommutative torus is the unital
$*$-algebra made of power series of the form \bbb a =\sum_{m,n \in
\IZ^2} a_{mn} ~U_1^m U_2^n~, \eee with $a_{mn}$ a complex-valued Schwarz function
on $\IZ^2$ that is, the sequence of complex numbers $\{a_{mn} \in \IC~, ~ (m,n) \in
\IZ^2 \}$ decreases rapidly at `infinity'. The two unitary elements $U_1, U_2$ have
commutation relations \bbb
\label{nct} U_2 ~U_1 = e^{2\pi i \theta} U_1 ~U_2~. \eee On $\ca_\theta$ there is a
unique normalized positive definite trace which we shall unusually denote by an
integral symbol $\int : \ca_\theta \raw \IC$ and which is given by \bbb
\int ( \sum_{(m,n) \in \IZ^2} a_{mn} ~U_1^m U_2^n) =: a_{00}~.\eee This trace is
invariant under the action of the commutative torus $T^2$ on $\ca_\theta$ whose
infinitesimal form is determined by two commuting derivations $\pa_1, \pa_2$ acting
by
\bbb\label{t2act}
\pa_\mu (U_\nu) = 2\pi i ~\delta_\mu^\nu U_\nu~, ~~\mu,\nu = 1, 2~. \eee The
invariance just being the statement that $\int \pa_\mu(a) = 0~, ~~ \mu = 1, 2~$ for
any $a\in\ca_\theta$.

All the previous properties, even if elementary, turn out to be important for
our construction since they allow us to use all tools of elementary calculus on a
commutative torus. However one must bear in mind that, in order to develop a more
general setting, one should work only with the corresponding cyclic and Hochschild
cocycles that we shall now describe.

The cyclic 2-cocycle associated to the integration of 2-forms is simply given by
\bbb
\psi(a_{0},a_{1},a_{2})={ i \over 2 \pi } \int \epsilon_{\mu\nu}
a_{0}\partial_{\mu}a_{1}\partial_{\nu}a_{2}, \eee where $\epsilon_{\mu\nu}$ is the
standard antisymmetric tensor. Its normalization ensures that for any hermitian
projector $p\in\aa_{\theta}$, the quantity $\psi(p,p,p)$ is an integer: it is indeed
the index of a Fredholm operator.

Working with the standard Euclidean metric on the torus, the Hochschild cocycle
$\phi$ is
\bbb\label{cocytorus}
\phi(a_{0},a_{1},a_{2})=
\frac{2}{\pi}\int a_{0}\partial a_{1}\ov{\partial}a_{2} \eee where the complex
derivations $\partial=1/2(\partial_{1}-i\partial_{2})$ and $\ov{\partial}=
1/2(\partial_{1}+i\partial_{2})$ are combination of the previous derivations. Note
that we consider $\partial$ and
$\ov{\partial}$ as maps with values in $\aa_{\theta}$ and not in the bimodule of
1-forms. A construction of the cocycle \eqn{cocytorus} as the conformal class of
the Euclidean metric on the torus can be found in \cite{book} and \cite{newconnes}.
We remark that one can also work with a general constant metric whose conformal
class is parametrized by a complex number $\tau$ that belongs to the upper half
plane. 

Accordingly, the action functional for our non-linear $\sigma$-model reads \bbb
\phi(1,p,p) = {1 \over 2 \pi} \int \partial_{\mu}p\partial_{\mu}p = {1 \over \pi}
\int p \partial_{\mu}p\partial_{\mu}p ~, \label{actfun} \eee the contraction with
the Euclidean metric being understood. 

As a subset of a topological vector space, the space $\pp_{\theta}$ of all
hermitian projectors of $\aa_{\theta}$ comes equipped with a natural topology (in
fact it is an infinite dimensional manifold) and we are interested in the study of
the critical points of the action \label{acfun} in a given connected component of
$\pp_{\theta}$. By carefully taking into account the non linear structure of the
space $\pp_{\theta}$, we get the field equations
\bbb
\label{eom} p ~\Delta(p) ~-~ \Delta(p) ~p = 0~. \eee where $\Delta = \pa_\mu
\pa_\mu$ is the laplacian. 

The previous equation is a non linear equation of the second order and it is rather
difficult to explicit its solutions in closed form. Following a standard route, we
shall show that the absolute minima of (\ref{acfun}) in a given connected component
of $\pp_{\theta}$ actually fulfill a first order equation which is easily solved. 

\par

Given a projector $p\in\pp_\theta$, there is a `topological charge' (the first
Chern number) defined by \cite{Coymt} \bbb \label{topcha} Q(p) =: {1 \over 2 \pi i}
\int p \Big[ \pa_1(p) \pa_2(p) - \pa_2(p)\pa_1(p) \Big] ~\in ~\IZ~.
\eee As in four dimensional Yang-Mills theory, this topological quantity yields a
bound for the action functional.

Due to positivity of the trace $\int$ and its cyclic property, we have
\bbb\label{pro} \int
\Big[ \pa_\mu(p) ~p \pm i \epsilon_{\mu\alpha} \pa_\alpha(p) ~p \Big]^* \Big[
\pa_\mu(p) ~p \pm i \epsilon_{\mu\beta} \pa_\beta(p) ~p \Big] \geq  0~, \eee from
which we obtain the inequality
\bbb\label{bpbou} S (p) \geq \pm 2 Q(p) ~.
\eee The inequality \eqn{bpbou}, which gives a lower bound for the action, is the
analogue of the one for ordinary $\sigma$-models \cite{BePo}. Also, it is a two
dimensional analogue of the inequality that occurs in four dimensional Yang-Mills
theory. A similar bound for a model on the fuzzy sphere has been obtained in
\cite{bala}.

{}From \eqn{pro} it is clear that the equality in \eqn{bpbou} occurs exactly when
the projector $p$ satisfies the following {\it self-duality} or {\it anti-self
duality} equations
\bbb
\label{sd0}
\Big[ \pa_\mu p \pm i \epsilon_{\mu\alpha} \pa_\alpha p \Big] ~p = 0 ~. \eee By
using the derivations $\pa , \bar{\pa} $ , the self duality equation \eqn{sd0}
reduce to
\bbb
\label{sd}
\bar{\pa} (p) ~p = 0~,
\eee while the anti-self duality one is
\bbb
\label{asd}
\pa (p) ~p = 0~.
\eee Simple manipulations show directly that each of the equations \eqn{sd} and
\eqn{asd} implies the field equations \eqn{eom}, as it should be. 

In the next section, we will partially solve these equations. 

\subsection{The instantons of charge $1$.} 

Before we proceed further, let us be more precise about the connected components of
$\pp_{\theta}$ \cite{canadian}. The latter are parametrized by two integers $m$ and
$n$ such that $m+n\theta>0$. When $\theta\in]0,1[$ is irrational, the corresponding
projectors are exactly the projectors of trace $m+n\theta$ and the topological
charge $Q(p)$ appearing in (\ref{topcha}) is just $n$. We shall construct our
solutions for $m=0$ and $n=1$ and postpone the general discussion to
\cite{non1}. Thus we have to find projectors that belongs to the previous homotopy
class and satisfy the self-duality equation $(\ov{\pa}p) p=0$ or, equivalently,
$p\pa p=0$.

Although these equations look very simple, they are far from being easy to solve
because of their non linear nature. To reduce them to a linear problem, we shall
introduce the following material and mimic the original construction of Rieffel
\cite{pacific} but with the constraint arising from the self-duality equation.

The space $\ee=\cs(\IR)$ of Schwarz functions of one variable is made into a {\it
right} module over $\ca_{-1/\theta}$ by defining \bbbb \label{rgtmod} && (\xi
~V_1)(s) =: \xi(s-1/\theta)~, \nn \\ && (\xi ~V_2)(s) =: e^{2\pi i s} \xi(s)~,
\eeee for any $\xi \in \ee$. It is easily checked that this defines an action on the
right of the algebra generated by $V_{1}$ and $V_{2}$ and that the latter is
isomorphic to $\aa_{-1 / \theta}$.

Furthermore, $\ee$ admits also a left action of $\aa_{\theta}$ given by \bbbb
\label{lftmod} && (U_1 \xi)(s) =: \xi(s-1)~, \nn \\ && (U_2 \xi)(s) =: e^{2\pi i
s\theta} \xi(s)~.
\eeee and one easily proves that the latter commutes with the right action of
$\aa_{-1/\theta}$. Besides, the elements of $\aa_{\theta}$ acting on the left are
exactly all linear operators from $\ee$ to itself that commute with the right
action of $\aa_{-1/\theta}$, namely $\aa_{\theta} \simeq {\rm
End}_{\aa_{-1/\theta}}(\ce)$.

On the module $\ee$ there is also a $\aa_{-1/\theta}$-valued hermitian structure,
namely a sesquilinear map (antilinear in the first variable) $\hs{~}{~} :
\ee\times\ee \raw\aa_{-1/\theta}$ which is compatible with the right
$\aa_{-1/\theta}$-module structure of $\ee$ (see \cite{book} for explicit formulae).
As a consequence, if $\xi\in\ee$ is such that $\hs{\xi}{\xi}$ is an invertible
element of $\aa_{-1/\theta}$, the endomorphism
\bbb p = \ket{\xi}{1 \over \hs{\xi}{\xi} }\bra{\xi} \eee is a self-adjoint
idempotent (that is a projector) in the algebra $\aa_{\theta}$ (due to the
identification
$\aa_{\theta} \simeq {\rm End}_{\aa_{-1/\theta}}(\ce)$). Here we are using a
physicist's notation for an element $\ket{\xi} \in \ce$ and the dual element
$\bra{\xi}
\in
\ce^*$ is defined by means of the hermitian structure as $\bra{\xi}(\eta) =
\hs{\xi}{\eta} \in \ca_{-1/\theta}$ for any $\eta \in \ce$. 

\bigskip

In order to translate the self-duality equations for $p$ to equations for $\xi$, we
need to introduce a connection on $\ce$. This is done \cite{CR} by means of two
covariant derivatives explicitly given by $\nabla_1,
\nabla_2:\,
\ee\rightarrow\ee$,
\bbb
\label{con} (\nabla_1 \xi)(s) =: {2\pi i\theta} s ~\xi(s)~, ~~~\nabla_2 \xi =: {d
\xi \over d s}~,
\eee These two operators fulfill a Leibniz rule with respect to the right action
\bbb
\nabla_\mu (\xi a) = (\nabla_\mu \xi)a + \xi (\pa_\mu a), ~~\mu=1, 2~. \eee for any
$\xi\in\ee$ and $a\in\aa_{-1/\theta}$ and $b\in\aa_{\theta}$. Furthermore, they are
compatible with the hermitian structure in the sense that \bbb
\pa_\mu \hs{\xi}{\eta} = \hs{\nabla_\mu \xi}{\eta} + \hs{\xi}{\nabla_\mu \eta}~,
~~\mu=1, 2~, \eee for any $\xi, \eta \in \ce$.

\bigskip

By introducing the operator $\ov{\nabla} = {1 /2} (\nabla_{1}+i\nabla_{2})$, it is
easy to show that
$p$ satisfies the self-dual equations \eqn{sd} if and only if there is an element
$\rho\in\aa_{-1/\theta}$ such that
\bbb
\label{nsd}
\ov{\nabla}\xi=\xi\rho.
\eee Thus, we manage to reduce the self-duality equation to a linear equation for
$\xi$ that can be easily solved in some simple cases. 

\par

When $\rho$ is a constant element (i.e. it is proportional to the unit of
$\aa_{-1/\theta}~, ~\rho=\lambda\, 1$, with $\lambda\in\ccc$), equation (\ref{nsd})
reduces to the simple differential equation \bbb
\frac{d\xi}{dt}+(2\pi\theta t+2i\lambda)\,\xi=0 \eee whose solutions are the
gaussians
\bbb\label{gauss}
\xi_{\lambda}(t)=A\,e^{-\pi\theta\,t^{2}-2i\lambda\,t}, \eee and $A\in\ccc^{*}$ is
an inessential normalization parameter.

\bigskip

We will show in \cite{non1} that, at least for $\theta$ small enough, the norms
$\hs{\xi_{\lambda}}{\xi_{\lambda}}$ are invertible. Accordingly, the gaussians
\eqn{gauss} provide a two (real) parameter family of solutions 
$p_\lambda = \ket{\xi_{\lambda}} 
( \hs{\xi_{\lambda}} {\xi_{\lambda}} )^{-1}
\bra{\xi_{\lambda}}$ of the self-duality equations
\eqn{sd}, and one can show that the freedom we have in $\lambda$ just corresponds
to the action of the ordinary torus on
$\aa_{\theta}$ by translation. Thus, we interpret the solution as a two dimensional
``instanton'' in this simple `` noncommutative Ising model'' (remember that the
target is just made of two points) and the freedom in $\lambda$ in a sense
corresponds to its location.

However, it is not obvious that different solutions of the self duality equations
on $\ee$, yield different projectors. In fact, $\xi$ and $\xi^{'}$ provide
different projectors if and only if they belong to different orbits of the action
of the group of invertible elements of $\aa_{-1/\theta}$ that acts on the right on
$\ee$. Obviously, this action preserves the invertibility of $\hs{\xi}{\xi}$ while
the structure of the self-duality equation (\ref{nsd}) is preserved provided $\rho$
is modified according to \bbb
\label{gauge}
\rho\rightarrow g^{-1}\rho g+g^{-1}\ov{\pa}g. \eee In a more physical language,
this means that in trading $p$ for $\xi$ we have introduced spurious gauge degrees
of freedom that we must get rid of. In the case of the Gaussians $\xi_{\lambda}$,
it is easy to show that $\xi_{\lambda}$ and $\xi_{\lambda^{'}}$ are gauge
equivalent if and only if
$\xi_{\lambda^{'}}=\xi_{\lambda}U_{1}^{n_{1}}U_{2}^{n_{2}}$ where $n_{1}$ and
$n_{2}$ are integers. More generally, given a solution $\xi$ of the self-dual
equation \bbb \ov{\nabla}\xi-\xi\rho=0
\eee with $\rho\in\aa_{-1/\theta}$, it is not clear that we can find a complex
gauge transformation $g$ (i.e. an invertible element of $\aa_{-1/\theta}$) that
allows to gauge transform $\xi$ into one of the gaussians. If this were the case,
it would mean that we had indeed constructed all self-dual solutions belonging to
the corresponding homotopy class. This problem is tantamount to solve the following
equation in $g$ and $\lambda$, \bbb
\rho=\lambda+g\ov{\pa}g^{-1}.
\eee Again, this can be done if $\theta$ is small enough \cite{non1}. The
corresponding idea is simple: we first notice that the problem is trivial when
$\theta=1/n$, with $n \in \nnn^{*}$ because $\aa_{-1/\theta}$ is commutative in
this case. Indeed, the existence of the gauge transformation results from the Hodge
decomposition of 1-forms. Then, we use the implicit function theorem in order to
find how to deform the commutative solution, considered as functions of $\theta$
\cite{non1}. 

A few additional remarks are in order. Even if many of the methods we have used are
similar to the ones used in the $CP^{N}$ model rather than to the ones pertaining
to the Ising model, we refrain from calling these ``noncommutative $CP^{N}$
models'' since we want to emphasize the fact that our target space in made of two
points and is not the manifold $CP^{N}$ (or more general grassmanian manifolds).
But obviously the ordinary grassmanian models can also be considered as
``noncommutative Ising models'' with a source described by matrix valued functions
over an ordinary Riemann surface.

It is also worth remarking that we have been working with the Euclidean metric, but
all constructions are readily extended to constant metrics whose conformal class
are parametrized by a complex number $\tau$ in the upper half-plane. Then, the
corresponding moduli space turns out to be a complex torus. 

\section{An analogue of the principal chiral model} 

Apart from finite spaces, the simplest possible target spaces are circles. Ordinary
two dimensional field theories compactified on a circle have been extensively
studied (see for instance \cite{gawedzki1}) and they essentially behave like free
fields (with minor deviations). As we shall show in our next example, this is not
the case for noncommutative models, the interaction arising from the noncommutative
nature of the source.

To proceed, let us first recall that the algebra of function over the circle $S^1$
is generated by a unitary element $U$. Thus, specifying a $*$-algebra morphism
$\pi$ from the algebra of functions on the circle to a another $*$-algebra $\aa$ is
tantamount to select a unitary element $g=\pi (U)$ in $\aa$. Accordingly, our
configuration space is made of all unitary elements of $\aa$. For the metric on the
circle we shall take the most natural one, $G=\delta U \delta U^{*}$, while for the
target space we take the Euclidean noncommutative torus, extension to other
constant metrics being straightforward. Then, the Hochschild cocycle is the one in
\eqn{cocytorus} and our action functional simply reads $S[g]=\phi(1,g,g^{-1})$,
which reduces to \bbb\label{prichimod} S[g]={1 \over 2 \pi}\int
\partial_{\mu}g\partial_{\mu}g^{-1}~, \eee the variables being unitary elements in
the algebra of the noncommutative torus
$\ca_\theta$.

Our model is analogous to a principal chiral model, with values in a unitary group
of matrices with which it shares lots of properties, apart from non-locality. For
the time being we shall limit our study to the existence of infinitely many
conserved currents.

{}From the action functional \eqn{prichimod}, one readily obtain the equations of
motion by varying $g\rightarrow g+\delta g$. As in the commutative case, they are
equivalent to a current conservation
\bbb
\partial_\mu\lp g^{-1}\partial_{\mu} g\rp=0, \eee which expresses the invariance
under the global $U(1)$ symmetry.

To construct infinitely many such currents, we use a standard induction that relies
on the Hodge decomposition of differential forms. Since the latter reduce to a
simple problem in linear algebra on the noncommutative torus we shall use it
without further discussion and write any differential form as a unique sum of a
harmonic one (i.e. a constant one), an exact one and a coexact one. 

Let us assume that we have constructed the conserved current $J_{\mu}^{(n)}$ and
let us build $J_{\mu}^{(n+1)}$. Since $J_{\mu}^{(n)}$ is conserved, the Hodge
decomposition just tells that it is a sum of a constant form and a co-exact one.
After an incorporation of the possible constant term into $J_{\mu}^{(n)}$, one can
find $\chi\in\aa_{\theta}$ such that \bbb
J_{\mu}^{(n)}=\epsilon_{\mu\nu}\partial_{\nu}\chi, \eee where $\epsilon_{\mu\nu}$
is the standard antisymmetric tensor. Then, let us introduce the gauge field
$A_{\mu}=g^{-1}\partial_{\mu}g$ and the covariant derivative
$D_{\mu}=\partial_{\mu}+A_{\mu}$. We define the next current as
\bbb J_{\mu}^{(n+1)}=D_{\mu}\chi~.
\eee It is easy to check that it is conserved, owing to the easily verified
commutation rules $[\partial_{\mu},D_{\mu}]=0$ and $[D_{\mu},D_{\nu}]=0$. Starting
with $J_{\mu}^{(1)}=g^{-1}\partial_{\mu}g$, by repeating the construction we can
construct an infinite number of non local conserved currents. 

Of course the series of new currents would stop whenever there appears a constant
current. With some more work one can show that this does not happen unless one
starts with a trivial solution of the equations of motion which is a product of the
generators. One could also object that non trivial solutions of the equations of
motion may not exist. Again this is not the case, since one can take $g=2p-1$,
where $p$ is one of the instantonic solutions we constructed in the previous
section.

All previous construction is very elementary and follows directly from the ordinary
field theoretical construction of the currents. The only point we want to emphasize
is that the latter still works in noncommutative geometry. A more thorough survey of
our theory along the classical lines \cite{karen} will be given in \cite{non2},
including a generalization of unitons.

\section{Addition of the Wess-Zumino term} 

Although the previous considerations are purely classical, the models can be
quantized. This amounts to define and compute the partition function
\bbb \zz =\int [\dd g] e^{-S[g]}~,
\eee together with the correlation functions
\bbb
\langle g\ot g\ot\cdots\ot g\rangle=\frac{1}{\zz} \int[\dd g](g\ot g\ot\cdots\ot g )
e^{-S[g]}, \eee as well as possible insertions of composite operators. Of course,
none of these functional integrals are well defined and to give a precise meaning
to them, one has to set up the renormalization procedure which yields some non
trivial problems even in dimension two, these models being power counting
renormalizable only in that dimension.

However, as far as the one-loop level is considered, this is easily achieved within
the background field method. As its non-abelian cousins, our model exhibits a
negative $\beta$ function so that one may say that it is asymptotically free.
Accordingly, one can definitely exclude the possibility of having a free field
theory.

We shall see that, after addition of the so called ``Wess-Zumino term'', the model
behaves almost like a free field (see, for instance, \cite{gawedzki2} and
\cite{shifman} for recent pedagogical reviews of the ordinary WZW model). Once
again, we will only be sketchy because of lack of space, and refer to \cite{non2}
for a detailed account.

To construct the Wess-Zumino term, let us start with a given unitary element $g$ of
$\aa_{\theta}$. It is known from K-theory \cite{canadian} that there always exist a
curve $g_{t}~, t\;\in [0,1]$ in the group of unitary elements of $\aa_{\theta}$
that fulfills $g_{1}=g$ and $g_{0}=(U_{1})^{n_{1}}(U_{2})^{n_{2}}$, where
$(n_{1},n_{2})$ denotes the class of
$g$ in K$_{1}(\aa_{\theta})$. Therefore, we can define the Wess-Zumino term as
\bbb\label{wzterm} S_{WZ}[g]=\frac{ik}{4\pi}\epsilon^{\mu\nu}\int_{0}^{1}dt\int
g^{-1}_{t}\frac{dg_t}{dt}\partial_{\mu}g_{t}^{-1}g_{t}, \eee where $k$ is an {\it a
priori} arbitrary real number. As in the classical case, this term can be expressed
as the integral over a solid noncommutative torus, but the latter depends on the
class of $g$ in K-theory, different classes yielding isomorphic and cobordant solid
tori.

Although the model \eqn{wzterm} depends on the curve $g_{t}$ and not only on $g$
one can show that, given any other curve $\tilde{g}_{t}$ connecting the same
boundaries, the difference of the two Wess-Zumino terms can be expressed as an
integral over a loop in the group of unitary elements of $\aa_{\theta}$. Such a
quantity may be easily identified with a coupling of a 3-cyclic cocycle with a
unitary element of $C^{\infty}(S^{1})\ot\aa_{\theta}$ and it can be shown to be
proportional to an integer, as follows from a straightforward application of the
index theorem (see \cite{the} for a very elementary treatment). It turns out that
if $k\in\zzz$, the Wess-Zumino term is defined up to integral multiples of
$2i\pi$.

Accordingly, we construct the Wess-Zumino-Witten action just by adding the previous
term to the non-linear $\sigma$-model and we get, \bbb
S_{WZW}[g]=\frac{k}{8\pi}\int\partial_{\mu}g\partial{\mu}g^{-1}
+\frac{ik}{4\pi}\epsilon^{\mu\nu}\int_{0}^{1}dt\int
g^{-1}_{t}\frac{dg_t}{dt}\partial_{\mu}g_{t}^{-1}g_{t}, \eee for positive $k$. 

By introducing the usual operators $\partial$ and $\ov{\partial}$, algebraic
manipulations involving integrations by part show that a Polyakov-Wiegman identity
holds, namely
\bbb S_{WZ}[gh]=S_{WZ}[g]+S_{WZ}[h]+
\frac{1}{4i\pi}\int g^{-1}\ov{\partial} g\,h\partial h^{-1}. \eee When 
$h=1+g^{-1}\delta
g$, this identity allows one to write the variation of $S_{WZW}[g]$ as \bbb \delta
S_{WZW}[g]=-\frac{k}{2i\pi}\int g^{-1}\delta g\partial\lp g^{-1}\ov{\partial}g\rp.
\eee 
Then, the equations of motion can be written equivalently as \bbb
\partial\ov{J}=0~, ~~~{\rm or}~~~ \ov{\partial}J=0~, \eee with
$\ov{J}=g^{-1}\ov{\partial}g$ and $J=g\partial g^{-1}$. 

One readily sees that there are very few solutions of the previous equation, since
any holomorphic function on the noncommutative torus, defined as an element of the
algebra in the kernel of $\ov{\partial}$, is constant. 

In order to get non trivial solutions, we equip the torus with the Minkowski metric.
Then the equations of motion are
\bbb\label{eommink}
\partial_{+}\lp g^{-1}\partial_{-}g \rp=0, \eee with
$\partial_{\pm}=\partial_{1}\mp\partial_{2}$. Apart from products of the generators
$U_{1}$ and $U_{2}$, we will show that the general solution of equations
\eqn{eommink} can be factorized as
\bbb g=g_{+}g_{-},
\eee where $g_{\pm}$ are unitary elements of $\ca_\theta$ satisfying the equations
$\partial_{\mp}g_{\pm}=0$.

To proceed, let us first assume that $g$ belongs to the connected component of the
identity. If this is not the case, we multiply it by a suitable product of the
generators, given by the class of $g$ in K-theory (the result is still a solution
of the equation of motion). Now, it follows from the equation \eqn{eommink} that
$J_{-}=g^{-1}\partial_{-}g$ belongs to the kernel of $\partial_{+}$ so that it can
be expanded as a Laurent series in $U_{1}U_{2}^{-1}$. Besides, since we are
assuming that $g$ belongs to the connected component of the identity, the constant
mode of the expansion vanishes, since it is invariant under deformation of $g$.
Therefore, the primitive
\bbb
\int_{-}\,J_{-}
\eee is well defined (one simply has to divide the coefficient in front of any
monomial by the corresponding non vanishing power). As a consequence, the solutions
of the remaining equation
\bbb
\partial_{-}g=gJ_{-}
\eee are easily expressed as
\bbb g=g_{+}e^{\int_{-}\,J_{-}}
\eee with $g_{+}$ an arbitrary unitary element of the algebra of Laurent series in
$U_{1}U_{2}$. Thus, $g_{\pm}$ can be expanded as \bbb
g_{\pm}=\mathop{\sum}\limits_{n\in\zzz} g^{(n)}_{\pm}\lp U_{1}U_{2}^{\pm 1}\rp^{n},
\eee and both can be interpreted as maps from circles $S^{1}_{\pm}$, which are the
spaces of characters of the commutative algebras generated by $U_{1}U_{2}^{\pm 1}$,
to $U(1)$. Note however that the coordinates on $S^1_{+}$ do not commute with the
ones on $S^1_{-}$.

Although this model almost looks like a free field theory, with commutative left and
right movers, the standard parity symmetry that exchanges left and right has been
broken and the theory, due to noncommutativity, always remembers that left movers
must appear on the left. Alternatively, this may be understood as a lack of
invariance of the Wess-Zumino term under the inversion $g\rightarrow g^{-1}$, while
the kinetic term obviously enjoys this symmetry. {}From the strict point of view of
solving the equation of motion, this is the only remainder of the noncommutative
nature of the source space.

Obviously, the space of solutions of the equations of motion is invariant under
gauge symmetry (respective multiplication on the left and the right by left and
right moving unitaries) and under conformal symmetry (reparametrisation of
$S^{\pm}$). However, general conformal transformations are not symmetries of the
noncommutative torus in the sense that they do not correspond to automorphisms of
the algebra $\aa_{\theta}$; it is only the translations that can be lifted to
automorphisms. Therefore, there is no {\it a priori} satisfactory way to define
conformal transformations of $g$ when it is not a solution of the equations of
motion. Note that gauge transformation do not create any trouble since they
correspond to inner automorphisms of the algebra.

Fortunately, one can construct analogues of conformal transformations that do leave
the action invariant and reduce both on-shell and in the commutative case to
ordinary conformal transformations. To proceed, let us introduce the (not
necessarily conserved) left and right currents $J_{\pm}$, analogous of $J$ and
$\ov{J}$. Furthermore, let us introduce infinitesimal multiplets
$\epsilon_{\pm}=(\epsilon_{\pm}^{(1)},\dots,\epsilon_{\pm}^{(n)})$ of left and
right moving elements of $\aa_{\theta}$. then, we define the infinitesimal
transformations
$\delta_{\epsilon_{\pm}}(g)$ as \bbb
\delta_{\epsilon_{-}}(g)=g\,\lp\mathop{\sum}\limits_{\mathrm{permutations}}
\epsilon_{-}^{(i_{1})}J_{-}\epsilon_{-}^{(i_{2})}J_{-}\dots
\epsilon_{-}^{(i_{n-1})}J_{-}\epsilon_{-}^{(i_{n})}\rp \eee and \bbb
\delta_{\epsilon_{+}}(g)=\lp\mathop{\sum}\limits_{\mathrm{permutations}}
\epsilon_{+}^{(i_{1})}J_{+}\epsilon_{+}^{(i_{2})}J_{+}\dots
\epsilon_{+}^{(i_{n-1})}J_{+}\epsilon_{+}^{(i_{n})}\rp\,g ~, \eee with the sums
running over all permutations of the indices $i_{1},\dots,i_{n}$. 

One readily sees that, by replacing $\delta g$ with $\delta_{\epsilon_{\pm}}(g)$ in
the variation of the Wess-Zumino-Witten action, one gets the integral of a total
derivative. Thus the variation vanishes even if $g$ is not a solution of the
equation of motion.

For the particular $n=1$ case, the previous transformations reduce to gauge
transformations. On shell or in the commutative case, the $n=2$ transformations are
just the conformal ones. The case $n>2$ is more exotic, since two such
transformations acting on the left and on the right do not in general commute.
Furthermore, it is not clear whether these transformations close off-shell or not.
Probably the closure of this algebra requires more general transformations. For
instance, $[\delta_{\epsilon_{+}},\delta_{\epsilon_{-}}]$ is a new symmetry that
one has to introduce into the algebra. In the same vein, one also introduces
transformations involving the derivatives of the currents. All these
transformations are of the form
\bbb
\delta_{\epsilon_{-},\epsilon_{+}}(g)=
g\,K_{-}(\epsilon_{-},J_{-})+K_{+}(\epsilon_{+},J_{+})\,g, \eee where
$K_{\pm}(\epsilon_{\pm},J_{\pm})$ are suitable products of the currents, their
derivatives and the corresponding parameters $\epsilon_{\pm}$. 

Whether this procedure ends or not is not so clear since any computation is rather
intricate due to the transformation of the currents themselves. One can also note
that the $n>2$ case will not yield symmetries of the ordinary $SU(N)$ non-abelian
Wess-Zumino-Witten theory (apart from the $SU(2)$ case), since it does not preserve
the unimodular condition of $SU(N)$, but they are bona fide transformations for
fields with values in $U(N)$.

\bigskip

As a final remark, we mention that we have constructed the Wess-Zumino term
associated with a particular cyclic cocycle of the
noncommutative torus. But the procedure is general and given any $2n$-cyclic
cocycle on an algebra
$\aa$ one can construct the associated Wess-Zumino term in a similar way.
Furthermore, the ambiguity in the definition will still be measured by the coupling
of a
$2n+1$-cyclic cocycle with an element of $K_1(C(S^{1})\ot\aa)$. This Wess-Zumino
term may be added to the action of a principal chiral field, constructed with a
Hochschild cocycle. In two dimensions, an analogue of the Polyakov-Wiegman identity
still holds provided one uses a suitable scalar product in the LHS. 

As a simple example, one can take the matrix algebra $\aa=M_{n}(\ccc)$ and the
cyclic cocycle given by the trace. Then, for any unitary matrix $g$, it is easy to
show that $S_{WZ}[g]=k\log\det g$. In this very simple case, the ambiguity in
defining the Wess-Zumino term is nothing but the ambiguity one encounters when
defining the argument of a complex number of modulus one, which is arbitrary up to
$2i\pi\zzz$. Furthermore, the Polyakov-Wiegman identity (we drop the kinetic term
in this example) reduces to the statement that the argument of a product is the sum
of the argument of its factors, up to $2i\pi\zzz$. 

\bigskip

We end here our sketchy discussion of the classical aspects of the noncommutative
Wess-Zumino-Witten model. We are aware that many interesting questions have been
left aside. In our opinion the main question is to understand how far from a free
field theory our model stands.

Finally, we mention that actions analogous to ours have been obtained
in \cite{shapo} and \cite{chu}. 

\bigskip\bigskip\bigskip
\bibliographystyle{unsrt}

\end{document}